\documentclass[notitlepage]{revtex4-1}

\usepackage{graphicx}
\usepackage{amsmath,url,multirow}
\usepackage{float} 

\begin{document}

\newenvironment{narrow}[2]{%
  \begin{list}{}{%
    \setlength{\topsep}{0pt}%
    \setlength{\leftmargin}{#1}%
    \setlength{\rightmargin}{#2}%
    \setlength{\listparindent}{\parindent}%
    \setlength{\itemindent}{\parindent}%
    \setlength{\parsep}{\parskip}%
  }%
  \item[]
}{\end{list}}

\newcommand{\installmentnum}{third}
\newcommand{\totalassets}{27}
\newcommand{\honeassets}{2}
\newcommand{\htwoassets}{25}
\newcommand{\futurepubdate}{2 May 2011}
\newcommand{\presentpubdate}{11 November 2010}

\title{
  The Financial Bubble Experiment: \\
  Advanced Diagnostics and Forecasts of Bubble Terminations \\
  Volume III--Master Document
}
\author{
  R. Woodard, D. Sornette, M. Fedorovsky \\
  (The Financial Crisis Observatory)
} 
\email{dsornette@ethz.ch}
\affiliation{
  Department of Management, Technology and Economics, ETH Zurich,
  Kreuzplatz 5, CH-8032 Zurich, Switzerland
}

\begin{abstract}
  This is the \installmentnum\ installment of the Financial Bubble
  Experiment. Here we provide the digital fingerprint of an electronic
  document~\cite{FBEv3_assets} in which we identify \totalassets\ bubbles in
  \totalassets\ different global assets; for \htwoassets\ of these assets, we
  present windows of dates of the most likely ending time of each bubble.  We
  will provide that document of the original analysis on \futurepubdate.
\end{abstract}

\date{\presentpubdate}

\maketitle

\begin{center}
  \textbf{\large UPDATE:  2 May 2011}
\end{center}

The names of the 27 assets can be found at the end of this document in
Section~\ref{sec:27-assets}.  The original assets document (whose checksum is
found in Table~\ref{table:checksums} in Section~\ref{sec:bubble-forecasts} of
this document) and the final analysis of the 27 assets can be found online at
\url{http://www.er.ethz.ch/fco/index}.  The remainder of this document,
excluding Section~\ref{sec:27-assets}, is the same as that uploaded on 12
November 2010 (and written on 11 November with data from 10 November) and can
be found as \texttt{v1} at \url{http://arxiv.org/abs/1011.2882}.  As always,
the reader is warmly encouraged to email the authors if any links cause trouble
and we will be happy to provide any documents requested.

\section{Introduction}
\label{Sec:Introduction}

The Financial Bubble Experiment (FBE) aims at testing the following two
hypotheses:
\begin{itemize}
\item {\bf Hypothesis H1:} Financial (and other) bubbles can be diagnosed in
  real-time before they end.
\item {\bf Hypothesis H2:} The termination of financial (and other) bubbles can
  be bracketed using probabilistic forecasts, with a reliability better than
  chance.
\end{itemize}
In a medical context, H1 corresponds to the diagnostic of cancer and H2 to the
forecast of remaining life expectancy.  

The motivation of the Financial Bubble Experiment finds its roots in the
failure of standard approaches. Indeed, neither the academic nor professional
literature provides a clear consensus for an operational definition of
financial bubbles or techniques for their diagnosis in real time.  Instead, the
literature reflects a permeating culture that simply assumes that any forecast
of a bubble's demise is inherently impossible.

Because back-testing is subjected to a host of possible biases, we propose the
FBE as a real-time advanced forecast methodology that is constructed to be
free, as much as possible, of all possible biases plaguing previous tests of
bubbles. In particular, active researchers are constantly tweaking their
procedures, so that predicted `events' become moving targets.  Only advance
forecasts can be free of data-snooping and other statistical biases of ex-post
tests.  The FBE aims at rigorously testing bubble predictability using methods
developed in our group and by other scholars over the last decade.  The main
concepts and techniques used for the FBE have been documented in numerous
papers
\cite{Jiang2010149,Johansen-Sornette-Ledoit-1999-JR,Johansen-Sornette-2006-BER,Sornette-Johansen-2001-QF,Sornette-Zhou-2006-IJF}
and the book \cite{Sornette-2003}.

In the FBE, we propose a new method of delivering our forecasts where the
results are revealed only after the predicted event has passed but where the
original date when we produced these same results can be publicly, digitally
authenticated.  Since our science and techniques involve forecasting, the best
test of a forecast is to publicize it and wait to see how accurate it is,
whether the wait involves days, weeks or months (we rarely make forecasts for
longer time scales).  We will do this and at the same time we want to delay the
unveiling of our results until after the forecasted event has passed to avoid
potential issues of liability, ethics and speculation.  Also, we think that a
full set of results showing multiple forecasts all at once is more revealing of
the quality of our current methods than would be a trickle of one such forecast
every month or so. We also want to address the obvious criticism of cherry
picking successful forecasts, as explained below. In order to be convincing,
our experiment has to report all cases, be they successes or failures.

The digital fingerprint of our first set of bubble forecasts was released on 2
November 2009 (with a hash update on 6 November 2009).  We added a new bubble
forecast on 23 December 2009.  The original forecasts and post-analysis were
presented publicly on 3 May 2010 and uploaded to the \texttt{arxiv} server on
14 May 2010.  All versions are available at~\cite{FBE-master-2009}.

This \installmentnum\ set of forecasts presents the methodology described
in~\cite{FBE-master-2009} and the digital fingerprint of a single document that
identifies and analyses \totalassets\ current asset bubbles (H1).  For
\htwoassets\ of those \totalassets\ bubbles, the document also provides windows
of dates of the most likely ending time of each bubble.  We will provide that
original document of the analysis on \futurepubdate.

\section{Description of the methodology of the 
 Financial Bubble Experiment}

Our method for this experiment is the following:

\begin{itemize}
\item We choose a series of dates with a fixed periodicity on which we will
  reveal our forecasts and make these dates public by immediately posting them
  on our University web site and on the first version of our main publication,
  which we describe below.  Specifically, our first publication of the
  forecasts was issued on 3 May 2010, with successive deliveries every 6
  months.  The forecasts of the current document will be presented on
  \futurepubdate.  However, we keep open the option of changing the periodicity
  of the future deliveries as the experiment unfolds and we learn from it and
  from feedback of the scientific community.
\item We then continue our current research involving analysis of thousands of
  global financial time series.
\item When we have a confident forecast, we summarize it in a simple
  \texttt{.pdf} document.
\item We do not make this document public.  Instead, we make its digital
  fingerprint public.  We generate two digital fingerprints for each document,
  with the publicly available 256 and 512 bit versions of the SHA-2 hash
  algorithm
  \footnote{\protect\url{http://en.wikipedia.org/wiki/SHA_hash_functions}}
  \footnote{ \protect\url{http://eprint.iacr.org/2008/270.pdf}}. This creates
  two strings of letters and numbers that are unique to this file.  Any change
  at all in the contents of this file will result in different SHA-2
  signatures.
\item We create the first version of our main document (this one), containing a
  brief description of our theory and methods, the SHA-2 hashes of our forecast
  and the date (\futurepubdate) on which we will make the original
  \texttt{.pdf} document public.
\item We upload this main `meta' document to \url{http://arxiv.org}.  This
  makes public our experiment and the SHA-2 hashes of our forecast. In
  addition, it generates an independent timestamp documenting the date on which
  we made (or at least uploaded) our forecast.  \url{arxiv.org} automatically
  places the date of when the document was first placed on its server as `v1'
  (version 1).  It is important for the integrity of the experiment that this
  date is documented by a trusted third party.
\item We continue our research until we find our next confident forecast.  We
  again put the forecast results in a \texttt{.pdf} document and generate the
  SHA-2 hashes.  We now update our master document with the date and digital
  fingerprint of this new forecast and upload this latest version of the master
  document to \url{arxiv.org}.  The server will call this `v2' (version 2) of
  the same document while keeping `v1' publicly available as a way to ensure
  integrity of the experiment (i.e., to ensure that we do not modify the SHA-2
  hashes in the original document).  Again, `v2' has a timestamp created by
  \url{arxiv.org}.
\item Notice that each new version contains the previous SHA-2 signatures, so
  that in the end there will be a list of dates of publication and associated
  SHA-2 signatures.
\item We continue this protocol until the future date (\futurepubdate) at
  which time we upload our final version of the master document.  For this
  final version, we include the URL of a web site where the \texttt{.pdf}
  documents of all of our past forecasts can be downloaded and independently
  checked for consistent SHA-2 hashes.  For convenience, we will include a
  summary of all of our forecasts in this final document.
\end{itemize}
Note that the above method implies two aspects of the same important check to
the integrity of our experiment:
\begin{enumerate}
\item We will reveal all forecasts, be they successful or not.
\item We will not simply `cherry-pick' the results that we would want the
  community to see (with a few token, possibly, bad results).  We do not have
  another simultaneous outlet where we are running a similar experiment, since
  \url{arxiv.org} is a very visible international platform.
\end{enumerate}

\section{Background and Theory}
\label{sec:background-theory}

Our theories of financial bubbles and crashes have been well-researched and
documented over the past 15 years in many papers and books.  We refer the
reader to the Bibliography.  In particular, broad overviews can be found in
\cite{Jiang2010149,Johansen-Sornette-Ledoit-1999-JR,Johansen-Sornette-2006-BER,Sornette-Johansen-2001-QF,Sornette-Zhou-2006-IJF}.
In short, our theories are based on positive feedback on the growth rate of an
asset's price by price, return and other financial and economic variables,
which lead to faster-than-exponential (power law) growth.  The positive
feedback is partially due to imitation and herding among humans, who are
actively trading the asset.  This signature is quantitatively identified in a
time series by a faster-than-exponential power law component, the existence of
increasing low-frequency volatility, these two ingredients occurring either in
isolation or simultaneously with varying relative amplitudes.  A convenient
representation has been found to be the existence of a power law growth
decorated by oscillations in the logarithm of time.  The simplest mathematical
embodiment is obtained as the first order expansion of the log-periodic power
law (LPPL) model and is shown in Eq.~\eqref{eq:lppl1}:
\begin{equation}
  \label{eq:lppl1} \ln P =  A + B |t - t_c|^\alpha + C |t - t_c|^\alpha
  \cos[\omega \ln |t - t_c| + \phi]
\end{equation}
where $P$ is the price of the asset and $t$ is time.  There are 7 parameters in
this nonlinear equation, whose relative importance and estimation are described
in our previous papers
\cite{Jiang2010149,Johansen-Sornette-Ledoit-1999-JR,Johansen-Sornette-2006-BER,Sornette-Johansen-2001-QF,Sornette-Zhou-2006-IJF}.
Our past work has led to the hypothesis that the LPPL signals can be useful
precursors to an ending (change of regime) of the bubble, either in a crash or
a less-dramatic leveling off of the growth.

\section{Methods}
\label{sec:methods}

\subsection{Bubble identification}
\label{sec:bubble-ident}

As are our theories, our methods are documented elsewhere so we only briefly
mention the general technique so that the forecasts that we make public can be
better understood.  In short, we scan thousands of financial time series each
week and identify regions in the series that are well-fit by
Eq.~\eqref{eq:lppl1}.  We divide each time series into sub-series defined by
start and end times, $t_1$ and $t_2$ and then fit each sub-series $(t_1, t_2)$.
We choose max($t_2$) as the date of the most recent available observation.
Many sub-series are created according to the following parameters: $dt_1 = dt_2
= 7$ days, min$(t_2 - t_1) = 91$ days and max$(t_2 - t_1) = 1092$ days.

After filtering all fits with an appropriate range of parameters, we select
those assets that have the strongest LPPL signatures.  To improve statistics,
we can calculate the residues between the model and the observations and use
the residues to create 10 synthetic datasets (bootstraps) that have similar
statistics as the original time series.  We fit Eq.~\eqref{eq:lppl1} to the
synthetic data and then extrapolate this entire ensemble of LPPL models to six
months beyond our last observation.  One of the parameters in the LPPL equation
is the ``crash'' time $t_c$, which represents the most probable time of the end
of the bubble and change of regime.  We identify the 20\%/80\% and 5\%/95\%
quantiles of $t_c$ of the fits of the ensemble consisting of original fits and
bootstrap fits.  These two sets of quantiles, the date of the last observation
and the number of fits in the ensemble are published in our forecasts.

\subsection{Post-analysis}
\label{sec:post-analysis}

Once the \texttt{.pdf} documents with the full description of the forecasts are
made public, the question arises as how to evaluate the quality of the
diagnostics and how these results help falsify the two hypotheses?  In a
nutshell, the problem boils down to qualifying (and quantifying) what is meant
by (i) a successful diagnostic of the existence of a bubble and (ii) a
successful forecast of the termination of the bubble.  In the end, one would
like to develop statistical tests to falsify the two hypotheses stated above,
using the track record that the present financial bubble experiment has the aim
to construct.  For instance, Chapter 9 of (Sornette, 2003) suggests a number of
options, including the ``statistical roulette'', Bayesian inference and error
diagrams. Our main goal with this FBE is to timestamp our forecasts as we
simultaneously continue our search for adequate measures to qualify the quality
of our forecasts.  

This quantification is an active, ongoing subset of our research.  We are
currently developing and testing novel estimations methods that will be
progressively implemented in future releases.  For our previous forecasts, we
quantified the quality of the forecasts with four measures that we will
continue to use in the final analysis:
\begin{itemize}
\item \textbf{Drawdown analysis:} Drawdown analysis simply identifies the
  largest drawdown observed between $t_2$ (date of forecast) and the date of
  the public `unveiling' of the original forecasts.  That is, we identify the
  largest drawdown in all available data after $t_2$.  A drawdown is simply
  defined as the largest peak-to-trough drop in price in a given region.
\item \textbf{Fraction of up days in a running window:} We calculate one day
  close-to-close returns for each asset and mark them as positive (up) or
  non-positive (zero or down).  The ratio of up days relative to the sum of up
  and down days in a running window of 30, 60 or 90 days is plotted on top of
  the price observations.
\item \textbf{Derivative of observations:} Another measure of the change of
  regime is provided by an estimation of the local growth rate.  We use the
  Savitzky-Golay smoothing algorithm to calculate the first derivative of the
  observations, using a third order polynomial fit centered within windows of
  120 and 180 days.
\item \textbf{Bubble index:} A measure we are developing that quantifies the
  quality of the LPPL fits to the price time series.
\end{itemize}
We are developing other measures that will be used in future analysis.

\section{Bubble Forecasts}
\label{sec:bubble-forecasts}

The checksums of the analysis document~\cite{FBEv3_assets} that contains the
names of the \totalassets\ assets are shown in Table~\ref{table:checksums}.
This document showing all \totalassets\ assets and \htwoassets\ forecasts, as
well as analysis of each identified bubble, will be uploaded to
\url{http://www.er.ethz.ch/fco/} on \futurepubdate.
\begin{table}[h]
{\tt
\begin{tabular}{|l|l|}
  \hline
  {Document name} &  \\
  \hline
  SHA256SUM & 4994beab18293be021d751d513b6fec0776fde9cf74c0098f7da8657487d950d \\
  \hline
  SHA512SUM & {\tiny ee20582b696a2ce880870b513e7b9e7ebb67bfbe62e2cad50dd18276a5158765af6fdf88d9fef6e047526c40478a865c722cab041386aa8efdd95da24dd9239d} \\
  \hline
\end{tabular}
}
\caption{Checksums of Financial Bubble Experiment Vol. III forecast document.}
\label{table:checksums}
\end{table}

\newpage

\section{The 27 Assets}
\label{sec:27-assets}

\begin{table}[H]
\begin{narrow}{3.5cm}{-1in}
  \begin{tabular}{|l|l|l|r|r|l|}
    \hline
    \multicolumn{5}{|c|}{\honeassets\ H1 Assets (identified bubble)} \\
    \hline
    Category & Asset & Ticker & H1 & C \\ 
    \hline
    Index & MERVAL Buenos Aires & \verb|^|MERV (Y) & 1 & * \\
    \hline
    Equity & AUTOZONE & AZO (Y) & 1 & \\
    \hline
  \end{tabular}
  \caption{\honeassets\ H1 assets of the Financial Bubble Experiment as of
    \presentpubdate.  All listed assets are candidates for H1 (identified
    bubble phase).  In the Ticker column, (B) stands for Bloomberg and (Y) for
    Yahoo Finance.  Columns `H1' and `H2' show a somewhat subjective score of
    -1 (worst), 0 or 1 (best), reflecting the quality of the forecasts.  This
    scoring is discussed further in Section~III of the main analysis document
    available at \texttt{http://www.er.ethz.ch/fco/index}.  Column `C' has an
    asterisk if an asset had a major correction within 3 days of $t_2=$2010-11-10
    (the last data observation used in our analysis). This correlated dynamics
    also is discussed in Section~III of the same analysis document.} 
\label{table:H1}
\end{narrow}
\end{table}

\begin{table}[H]
\begin{narrow}{-.9cm}{-1in}
  \begin{tabular}{|l|l|l|l|l|r|r|l|}
    \hline
    \multicolumn{8}{|c|}{\htwoassets\ H1 and H2 Assets (identified bubble)} \\
    \hline
    Category & Asset & Ticker & $t_c$ 20\% - 80\% & $t_c$ 5\% - 95\% & H1 & H2 & C \\ 
    \hline
    \multirow{12}{*}{Index} 
    & BSE SENSEX, Bombay & \verb|^|BSESN (Y) & 2010-11-03 - 2010-12-01 & 2010-10-27 - 2010-12-10 & 0 & 1 & * \\
    & Dow Jones-AIG Comm. & \verb|^|DJC (Y) & 2010-11-16 - 2010-12-04 & 2010-11-09 - 2010-12-10 & 1 & 1 & * \\
    & FTSE 100 & \verb|^|FTSE (Y) & 2010-11-27 - 2010-12-26 & 2010-11-07 - 2011-01-03 & 1 & 1 & * \\
    & Hang Seng Index & \verb|^|HSI (Y) & 2010-11-09 - 2010-12-09 & 2010-11-07 - 2010-12-16 & 1 & 1 & * \\
    & Interactive Week Internet & \verb|^|IIX (Y) & 2010-11-12 - 2010-12-10 & 2010-11-04 - 2010-12-23 & 1 & 1 & * \\
    & NASDAQ Computer & \verb|^|IXK (Y) & 2010-11-13 - 2010-12-06 & 2010-11-07 - 2010-12-09 & 1 & 1  & * \\
    & Jakarta Composite & \verb|^|JKSE (Y) & 2010-11-06 - 2010-12-09 & 2010-10-23 - 2010-12-25 & 0 & 0 &  \\
    & KOSPI Composite Index & \verb|^|KS11 (Y) & 2010-11-15 - 2010-12-26 & 2010-10-30 - 2011-01-07 & 1 & 0 & * \\
    & NASDAQ-100 (DRM) & \verb|^|NDX (Y) & 2010-11-05 - 2010-11-29 & 2010-11-03 - 2010-12-22 & 1 & 1 & * \\
    & Reuters/Jefferies CRB & CRY INDEX (B) & 2010-11-11 - 2010-11-22 & 2010-11-07 - 2010-11-26 & 1 & 1 & * \\
    & TSEC weighted index & \verb|^|TWII (Y) & 2010-12-01 - 2011-01-03 & 2010-11-13 - 2011-01-08 & 1 & -1 & * \\
    & Major Market Index & \verb|^|XMI (Y) & 2010-11-10 - 2010-11-25 & 2010-10-30 - 2010-12-04 & 1 & 1 & * \\
    \hline
    \multirow{6}{*}{Equity} 
    & Ishares Singapore Index & EWS (Y) & 2010-11-14 - 2010-12-12 & 2010-11-06 - 2010-12-25 & 1 & 1 & * \\
    & Freeport McMoRan & FCX (Y) & 2010-11-15 - 2010-12-17 & 2010-11-09 - 2010-12-27 & 1 & 1 & * \\
    & F5 NETWORKS & FFIV (Y) & 2010-12-27 - 2011-03-09 & 2010-12-02 - 2011-04-08 & 1 & 1 &  \\
    & INTUIT & INTU (Y) & 2010-11-28 - 2011-01-15 & 2010-11-07 - 2011-02-11 & 0 & 0 & * \\
    & STARBUCKS & SBUX (Y) & 2010-11-08 - 2010-11-18 & 2010-11-06 - 2010-11-25 & 1 & -1 & \\
    & UNITED RENTALS INC & URI (Y) & 2010-11-09 - 2010-12-13 & 2010-11-02 - 2011-01-08 & 1 & -1 &  \\
    \hline
    \multirow{6}{*}{Commodity}
    & Copper future (USD) & HG1 COMB Comdty (B) & 2010-11-09 - 2011-01-07 & 2010-10-31 - 2011-01-15 & 1 & 1 & * \\
    & Corn future (CHF) & C 1 COMB Comdty (B) & 2010-11-18 - 2010-12-19 & 2010-11-08 - 2010-12-28 & 1 & 1 & * \\
    & Cotton future (USD) & CT1 COMB Comdty (B) & 2010-11-12 - 2010-11-13 & 2010-11-08 - 2010-11-15 & 1 & 1 & * \\
    & Palladium future (USD) & PA1 COMB Comdty (B) & 2010-11-12 - 2010-11-19 & 2010-11-10 - 2010-11-27 & 1 & 0 & * \\
    & Silver future (CHF) & SI1 COMB Comdty (B) & 2010-11-13 - 2010-11-18 & 2010-11-08 - 2010-11-29 & 1 & 0 & * \\
    & Sugar future (CHF) & SB1 COMB Comdty (B) & 2010-11-20 - 2010-12-09 & 2010-11-10 - 2010-12-17 & 1 & 1 & * \\
    \hline
    Forex
    & AUDUSD &  (B) & 2010-11-12 - 2010-12-25 & 2010-10-30 - 2011-01-12 & 1 & 1 & * \\
    \hline
  \end{tabular}
  \caption{\htwoassets\ H2 assets of the Financial Bubble Experiment as of
    \presentpubdate.  All listed assets are candidates for H1 (identified
    bubble phase) and H2 (identification of end of bubble phase). Quantile
    windows of most likely dates of the end of the bubble phases are shown.
    Abbreviations (B), (Y), H1, H2 and C are described in caption of
    Table~\ref{table:H1}.} 
  \label{table:H2}
\end{narrow}
\end{table}

\newpage

\bibliography{fbe}

\end{document}